\documentclass[prl,twocolumn,showpacs,amsmath,amssymb]{revtex4}
\usepackage{graphicx}
\usepackage{dcolumn}
\usepackage{epsfig}

\begin{document}

\title{QCD Dirac operator at nonzero chemical potential: lattice data
  and matrix model}

\author{Gernot Akemann}
\affiliation{
  Service de Physique Th\'eorique, CEA/DSM/SPhT Saclay,\\
  Unit\'e de recherche associ\'ee au CNRS,
  F-91191 Gif-sur-Yvette Cedex, France}

\author{Tilo Wettig}
\affiliation{
  Department of Physics, Yale University, New Haven, CT 06520-8120,
  USA\\
  and RIKEN-BNL Research Center, Brookhaven National Laboratory,
  Upton, NY 11973-5000, USA}

\received{5 August 2003}

\begin{abstract}
  Recently, a non-Hermitian chiral random matrix model was proposed to
  describe the eigenvalues of the QCD Dirac operator at nonzero
  chemical potential.  This matrix model can be constructed from
  QCD by mapping it to an equivalent matrix model which has the same
  symmetries as QCD with chemical potential. Its microscopic spectral
  correlations are conjectured to be identical to those of the QCD
  Dirac operator.  We investigate this conjecture by comparing large
  ensembles of Dirac eigenvalues in quenched SU(3) lattice QCD at
  nonzero chemical potential to the analytical predictions of the
  matrix model.  Excellent agreement is found in the two regimes of
  weak and strong non-Hermiticity, for several different lattice
  volumes.
\end{abstract}



\maketitle

There has been a lot of recent interest in physical systems described
by non-Hermitian operators.  Such operators play a r\^ole in many areas
of physics, e.g., S-matrix theory \cite{VWZ85}, dissipative quantum
maps \cite{GHS88}, neural network dynamics \cite{LS91}, disordered
systems with imaginary vector potential \cite{HN96}, and quantum
chromodynamics (QCD) at nonzero density \cite{Step96}.  In the present
work, we are mainly interested in the last of these applications, but
we expect the matrix model to be described below to be applicable to
non-Hermitian operators in other physical systems as well, provided
they are in the same symmetry class.

QCD at nonzero density is important in a variety of physical
situations, such as relativistic heavy-ion collisions or neutron
stars.  Considerable progress has been made in the last few years on
the analytical side.  For example, the regime of asymptotically large
density is well understood \cite{ARW98}, and qualitative predictions
for the QCD phase diagram could be derived on the basis of symmetry
considerations \cite{HJSSV98}.  
However, quantitative results at physically relevant densities 
are still lacking at present.
Unfortunately, lattice simulations of full QCD at nonzero chemical
potential $\mu$ are extremely difficult: the weight function is
complex, and the numerical effort increases exponentially with the
volume.  A number of interesting new ideas have recently been
investigated on the lattice side, e.g., reweighting along the critical
line \cite{FK02}, combined expansions of weight function and
observable \cite{BielSwan02}, analytic continuation from imaginary
$\mu$ \cite{dFP02}, and a factorization method for distribution
functions of observables \cite{AANV02}.  It is questionable, however,
whether the techniques using real $\mu$ will allow us to approach 
the thermodynamic limit.

Clearly, a better theoretical understanding of QCD at nonzero density
is desirable.  The Dirac operator is one of the central objects in
QCD.  Many observables can be expressed in terms of its eigenvalues
and eigenvectors.  While much is known about this eigenvalue spectrum
at $\mu=0$ (see Ref.~\cite{VW00} for a review), the situation at
$\mu\ne0$ is less satisfying.  The goal of the present work is to
improve our understanding of the latter case.  We concentrate on a
particular matrix model for the QCD Dirac operator at nonzero $\mu$
and show that its analytical predictions for the distribution of small
Dirac eigenvalues are in agreement with data from lattice gauge
simulations.  This statement holds in the two different regimes of
weak and strong non-Hermiticity, to be defined below.  The implications
of these results are discussed in the conclusions.

We start by presenting the matrix model and its predictions
\cite{A02}.  The model constitutes a complex extension of the chiral
Gaussian Unitary Ensemble (GUE) \cite{SV93}.  In terms of the complex
eigenvalues $z_j$ ($j=1,\ldots,N$), its partition function reads
\begin{align}
  \label{eq:Zev}
  Z_\nu(\tau;\{m_f\}) &= \int\limits_{\mathbb{C}}\prod_{j=1}^N
  dz_j dz_j^\ast \ |z_j|^{2|\nu|+1}\prod_{f=1}^{N_f}(z_j^2+m_f^2)\notag\\[-1mm]
  &\times \mbox{e}^{-\frac{N}{1-\tau^2}\left[|z_j|^2
  -\frac{\tau}{2}(z_j^2+z_j^{\ast\,2})\right]}
\prod_{k>l}^N \left|z_k^2-z_l^2\right|^2\!
\end{align}
for $N_f$ flavors of masses $m_f$ ($f\!=\!1,\ldots,N_f$) in the sec\-tor
of topological charge $\nu$.  The parameter $\tau\!\in\![0,1]$
measures the degree of non-Hermiticity and is related to $\mu$ by
\begin{equation}
\label{mutau}
  \mu^2=1-\tau^2\:.
\end{equation}
In the limit $\tau\to1$ (or $\mu\to0$), the
eigenvalues are real and we are back to the chiral GUE.
For $\tau\to0$ the non-Hermiticity is maximal, and the model becomes a
chiral extension of the Ginibre Ensemble \cite{Gini65}.  
The relation \eqref{mutau} follows 
from comparing the current model to the matrix model of Ref.~\cite{Step96} 
at small $\mu$.  
That model has the same global symmetries as QCD and is defined by  
\begin{align}
\label{Zsteph}
\hat Z_\nu(\mu;\{m_f\}) = \!\!\!\int\limits_{\mathbb{C}^{(N+\nu)\times
    N}}\!\!\!\!\! d\Phi \:\prod^{ N_f}_{f=1} 
&\det\left[
\begin{array}{cc}
m_f                 & i\Phi + \mu\\
i\Phi^\dagger + \mu &         m_f\\
\end{array}
\right]\notag\\[-1mm]
 \times &\exp[-\frac{N}2\,\mbox{Tr}\,\Phi^\dagger\Phi]\ .
\end{align}
The current model
\eqref{eq:Zev} is equivalent to the model \eqref{Zsteph} at the level
of the partition function for small values of $\mu$ \cite{A03}.
(However, unlike the model \eqref{Zsteph}, it is always in the phase
with broken chiral symmetry.)  
So far it has only been possible to compute
the microscopic spectral correlations (i.e., the correlations of the
smallest eigenvalues on the scale of the mean level spacing) for model
\eqref{eq:Zev} \cite{A02}, and not for model \eqref{Zsteph}.  
It is conjectured that the two
models, as well as QCD, are in the same universality class in the
sense that they yield identical results for the microscopic spectral
correlations.
The 
analytical predictions for model \eqref{eq:Zev} 
can be derived at large $N$ using the technique of orthogonal
polynomials in the complex plane \cite{A02}. 
All correlation functions have been
obtained either for $N_f=0$, or for $N_f\neq0$ ``phase-quenched'' massless
flavors, replacing $z_j^{2N_f}\to|z_j|^{2N_f}$.

Two different large-$N$ limits have to
be distinguished. In the limit of weak non-Hermiticity \cite{FKS97},
the product 
\begin{equation}
\label{weaklim}
  \lim_{N\to\infty}\lim_{\tau\to1}\ N(1-\tau^2) =
\lim_{N\to\infty}\lim_{\mu\to0}\ N\mu^2 
\equiv \alpha^2 \ .
\end{equation}
is kept fixed. 
This corresponds to taking the volume $V\propto N$ to infinity
such that $V\mu^2$ is fixed.  The result of the current model for the
density $\rho(z)=\langle\prod_{j=1}^N\delta(z-z_j)\rangle$  
of small Dirac eigenvalues in this limit reads at $N_f=0$
(corresponding to the quenched limit in our lattice data)
\begin{align}
\label{rhoweak}
  \rho_{\rm weak}(\xi) =\ & \frac{\sqrt{\pi\alpha^2}}
  {\mbox{erf}(\alpha)}\ |\xi|\ \exp\!\left[-\frac{(\Im
    m\,\xi)^2}{\alpha^2}\right] \notag\\ 
  & \times \int_0^1 dt\ \mbox{e}^{-\alpha^2t}
  J_{|\nu|}(\sqrt{t}\xi)J_{|\nu|}(\sqrt{t}\xi^*)\:,
\end{align}
where $J$ denotes the Bessel function and the eigenvalues have been
rescaled according to $\xi=\sqrt{2}Nz$, resulting in the customary
level spacing of $\pi$.
The limit of strong non-Hermiticity, with $N\to\infty$ at fixed
$\tau\!\in\![0,1)$, 
leads to
\begin{equation}
\label{rhostrong}
  \rho_{\rm strong}(\xi) = \sqrt{2\pi}\,
   |\xi| \exp\!\left(-|\xi|^2\right) \!
I_{|\nu|}\!\left(|\xi|^2\right),\!\!
\end{equation}
where $I$ denotes the modified Bessel function.  The rescaling
$\xi\!=\!\sqrt{N/(1\!-\!\tau^2)}\:z$ results in a level spacing
independent of $\tau$.  We stress that the existence of these two
different scaling regimes is a prediction for the lattice, and we will
identify these two regimes in the data below.

We now turn to the details of the lattice simulations and discuss some
of the concerns arising from our choices of operator and simulation
parameters.  We use the staggered Dirac operator, given at $\mu\ne0$ 
in terms of SU(3) gauge fields $U$ and staggered phases $\eta$ 
by \cite{HK83}
\begin{align}
  \label{Dirac}
  D_{x,y}(U,\mu) =\; &
  \frac12 \sum\limits_{\nu=\hat{x},\hat{y},\hat{z}}
  \left[U_{\nu}(x)\eta_{\nu}(x)\delta_{y,x\!+\!\nu}-{\rm h.c.}\right]
  \notag\\
  &\hspace{-18mm}
  +\frac12\left[U_{\hat{t}}(x)\eta_{\hat{t}}(x)e^{\mu}
    \delta_{y,x\!+\!\hat{t}}
    -U_{\hat{t}}^{\dagger}(y)\eta_{\hat{t}}(y)
    e^{-\mu}\delta_{y,x\!-\!\hat{t}}\right]\:.
\end{align}
The lattice spacing has been set to unity. 
We denote its eigenvalues by $i\lambda_k$ with $\lambda_k$ real (complex)
for $\mu=0$ ($\mu\neq0$). The reason to prefer the staggered formulation
is that (a) the Wilson operator 
breaks chiral symmetry explicitly and
has complex eigenvalues even at $\mu=0$
and 
(b) Ginsparg-Wilson-type operators 
are much more
expensive to compute. This is a serious issue here because we need
many configurations (see Table~\ref{table:par}).

The gauge field configurations were generated in the quenched
approximation, corresponding to $N_f=0$.  At $\mu\ne0$ the $N_f\to0$
limit is subtle.  In the current model, this limit can be taken in
different ways: (i) use Eq.~\eqref{eq:Zev} and let $N_f\to0$ at the
end of the calculation, (ii) use Eq.~\eqref{eq:Zev}
``phase-quenched'', i.e.\ with $|z_j|^{2N_f}$ in the weight, and let
$N_f\to0$ at the end of the calculation, or (iii) set $N_f=0$ in
Eq.~\eqref{eq:Zev} at the beginning of the calculation.  (ii) and
(iii) yield identical results, given in
Eqs.~(\ref{rhoweak},\ref{rhostrong}), and these results agree with
quenched lattice data at $\mu\ne0$, as shown below.  This is
consistent with the fact that quenched QCD at $\mu\ne0$ is the
$N_f\to0$ limit of a theory in which $(\det D)^{N_f}$ is replaced by
$|\det D|^{N_f}$ \cite{Step96}.  Taking the $N_f\to0$ limit according
to (i) will lead to a different (yet unknown) result which would not
describe the quenched lattice data.  At $N_f>0$, however, we expect
Eq.~\eqref{eq:Zev} to be valid for unquenched QCD, since it contains
the correct phase of the fermion determinant.  This expectation is
strengthened by the fact that matrix models of this type have already
been successful in describing unquenched QCD at $\mu\ne0$
\cite{HJSSV98,AANV02}.  (We also note that matrix models describe
unquenched lattice data at $\mu=0$ \cite{BMW98}.)

\begin{table}[b]
  \vspace*{-3mm}
  \caption{Summary of simulation parameters ($\beta=5.0$).}
  \label{table:par}
  \begin{ruledtabular}
    \begin{tabular}{rdcc}
      \multicolumn{1}{c}{$V$} & \multicolumn{1}{c}{$\;\;\;\;\;\mu$} &
      level spacing $d$ & no. of config. \\[1mm]
      $6^4$ & 0.006   & $1.98(2) \cdot 10^{-3}$ & 17,000 \\
      $6^4$ & 0.03    & $1.55(3) \cdot 10^{-3}$ & 20,000 \\
      $6^4$ & 0.2     & $6.83(6) \cdot 10^{-3}$ & 20,000 \\
      $8^4$ & 0.003375& $6.30(3) \cdot 10^{-4}$ & 20,000 \\
      $8^4$ & 0.2     & $3.85(3) \cdot 10^{-3}$ & 20,000 \\
      $10^4$ & 0.00216& $2.57(4) \cdot 10^{-4}$ & 4,000\\
      $10^4$ & 0.2    & $2.46(4) \cdot 10^{-3}$ & 4,000
    \end{tabular}
  \end{ruledtabular}
\end{table}

In the simulations we used $\beta=6/g^2=5.0$, which corresponds to the
strong-coupling regime and is far from the continuum limit. 
Nevertheless, working at such a low
value of $\beta$ is both convenient and legitimate for the present
purpose, for the following reason.  The microscopic spectral
correlations are described by a random matrix model only below the
so-called Thouless energy, which is the boundary of the regime in
which the zero-momentum modes dominate the partition function of the
low-energy effective theory in a finite volume \cite{VW00}.  The
Thouless energy is a function of both $\beta$ and the lattice volume
$V$.  If $\beta$ is increased at fixed $V$, fewer Dirac eigenvalues
are described by the matrix model.  Increasing $V$ works in the
opposite direction.  Thus, for small $\beta$ we can test the matrix
model on relatively small lattices.  At larger values of $\beta$ we
simply need to increase the lattice volume, which, however, is
inconvenient numerically.

At this small value of $\beta$, the staggered Dirac operator does not
have exact zero modes even if the underlying gauge field has nonzero
topological charge, because the would-be zero modes are shifted by an
amount proportional to the square of the lattice spacing \cite{SV88}.
This amount is much larger than the mean level spacing near zero, and
thus the would-be zero modes are completely mixed with the nonzero
modes.  We account for this by setting $\nu=0$ in
Eqs.~(\ref{eq:Zev},\ref{rhoweak},\ref{rhostrong}).  (This ``disease'' of
staggered fermions can be overcome by going to very small lattice
spacing \cite{FHL99} or by using the overlap operator \cite{EHKN99}.)

Our simulation
parameters are summarized in Table~\ref{table:par}.  
In the regime of strong non-Hermiticity, we used a constant value of
$\mu=0.2$, whereas for weak non-Hermiticity, we varied $\mu$ such that
the product $\mu^2V$ is fixed.  The gauge fields were generated using
a combined overrelaxation and Metropolis algorithm, written by
P.E.L. Rakow. 
On the $6^4$ lattice, the complete
eigenvalue spectrum can be computed on a single PC rather quickly
using LAPACK \cite{LAPACK}.  For the larger lattices we switched to
ARPACK \cite{ARPACK} and computed only the 100 eigenvalues of smallest
absolute magnitude with positive real part (the eigenvalues 
come in pairs $\pm\lambda_k$).  Since ARPACK is very fast at computing
the largest eigenvalues, we inverted the Dirac operator prior to
feeding it to ARPACK, using the sparse LU solver UMFPACK
\cite{UMFPACK}.

Because we are only interested in the small eigenvalues of the Dirac
operator, global unfolding of the spectrum \cite{MPW99} is not
necessary; we simply rescale the eigenvalues by a constant determined
from the mean level spacing $d$ of the data near zero.  Note that 
$d\propto1/V$ in the weak limit and $d\propto1/\sqrt{V}$ in the strong
limit, respectively.
 
\begin{figure}[h]
\centerline{
  \epsfig{figure=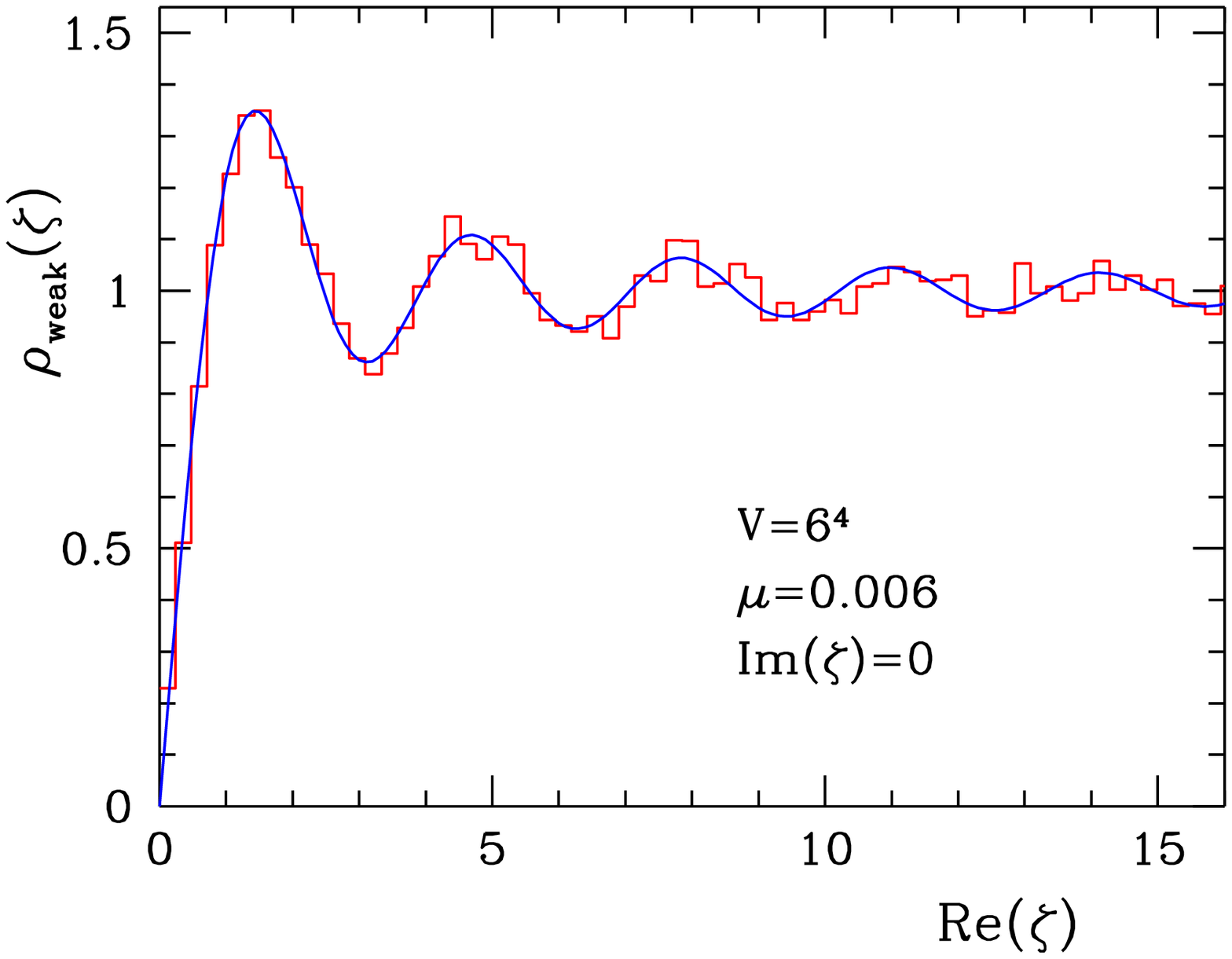,height=33mm}\hspace*{4mm}
  \epsfig{figure=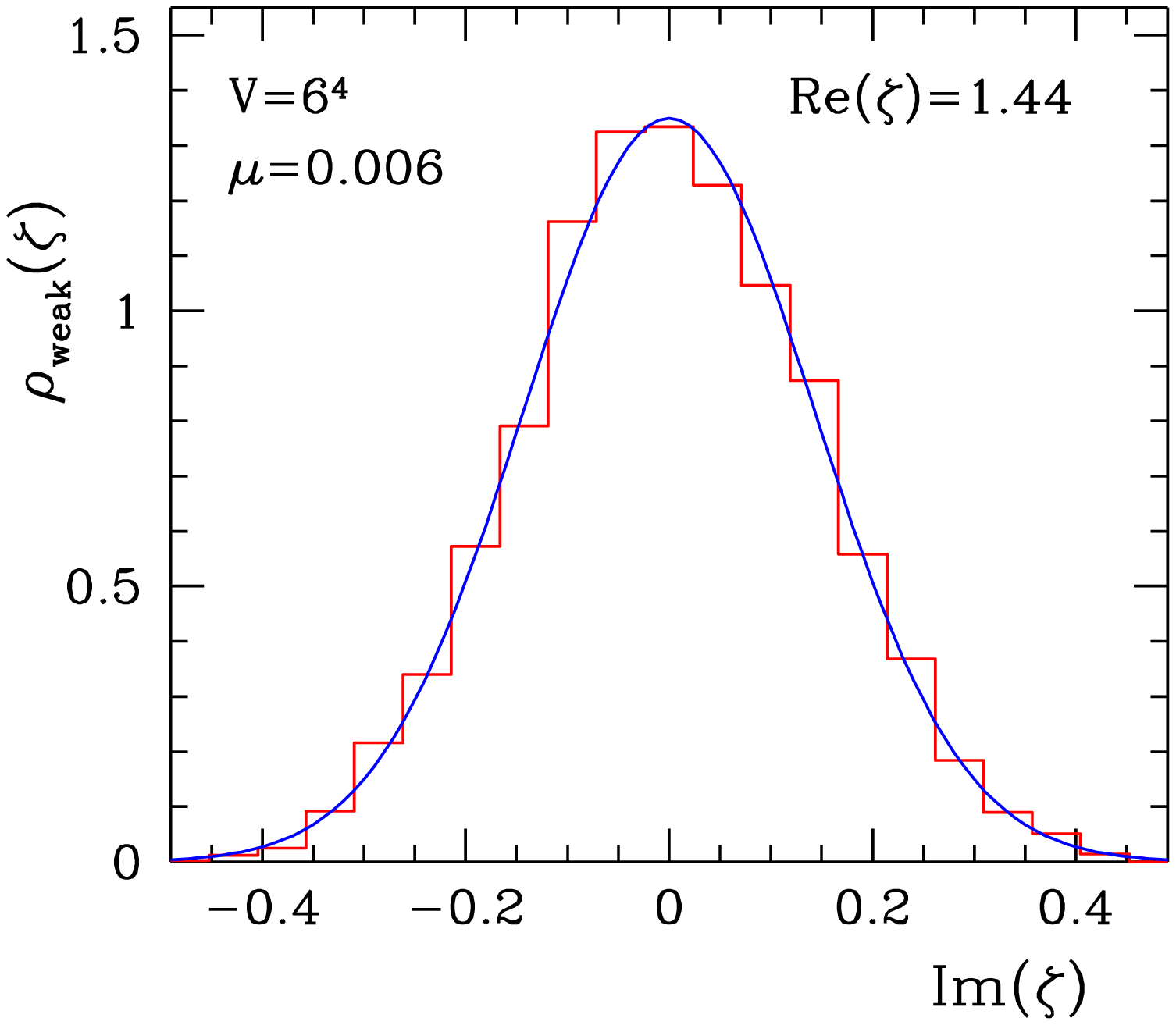,height=33mm}
}
\caption{
  Density of small Dirac eigenvalues for $V=6^4$ and $\mu=0.006$, cut
  along the real axis (left) and parallel to the imaginary axis at the
  first maximum (right).  The histogram represents lattice data, and
  the solid curve is Eq.~\eqref{rhoweak}.}
\label{weak6^4}
\end{figure}

We first present our results for weak non-Hermiticity.  Since
three-dimensional plots are hard to read, we instead show cuts along
the real and imaginary axes.  In Fig.~\ref{weak6^4} the data for
lattice size $6^4$ and $\mu=0.006$ are plotted versus
Eq.~\eqref{rhoweak}.
There is no free fit parameter; the data have been rescaled according
to $\xi=\pi \lambda/d$, with $d\propto1/V$ given in Table~\ref{table:par}
(note that $d$ depends on the lattice spacing, i.e.\ it would change
with $\beta$).   
At weak non-Hermiticity $d$ can be obtained in the same way as for real
eigenvalues: Because of the smallness of their imaginary part the
eigenvalues can still be ordered with respect to their real part so
that the level spacing is defined unambiguously.  The very same level
spacing $d$ is used to determine $\alpha^2=(\pi/\sqrt{2})\,\mu^2/d$
from Eq.~\eqref{weaklim} and $d_{\rm model}=\pi/(\sqrt{2}N)$, leading
to $\alpha=0.20$ for use in Eq.~\eqref{rhoweak}. Within our
statistical accuracy, we obtain excellent agreement of lattice data
and analytical prediction without any free fit parameter.
\begin{figure}[b]
  \centerline{ \epsfig{figure=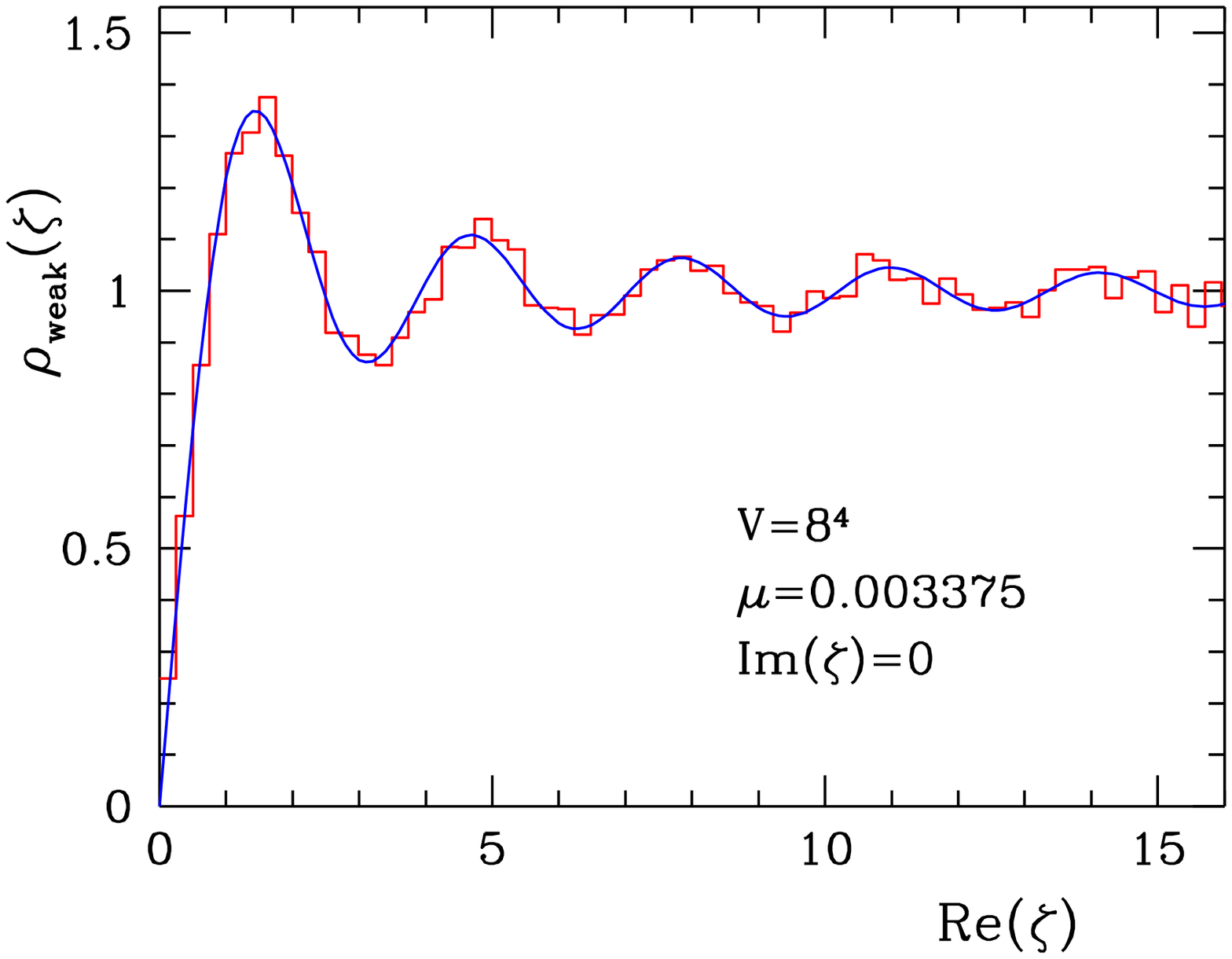,height=33mm}\hspace*{4mm}
    \epsfig{figure=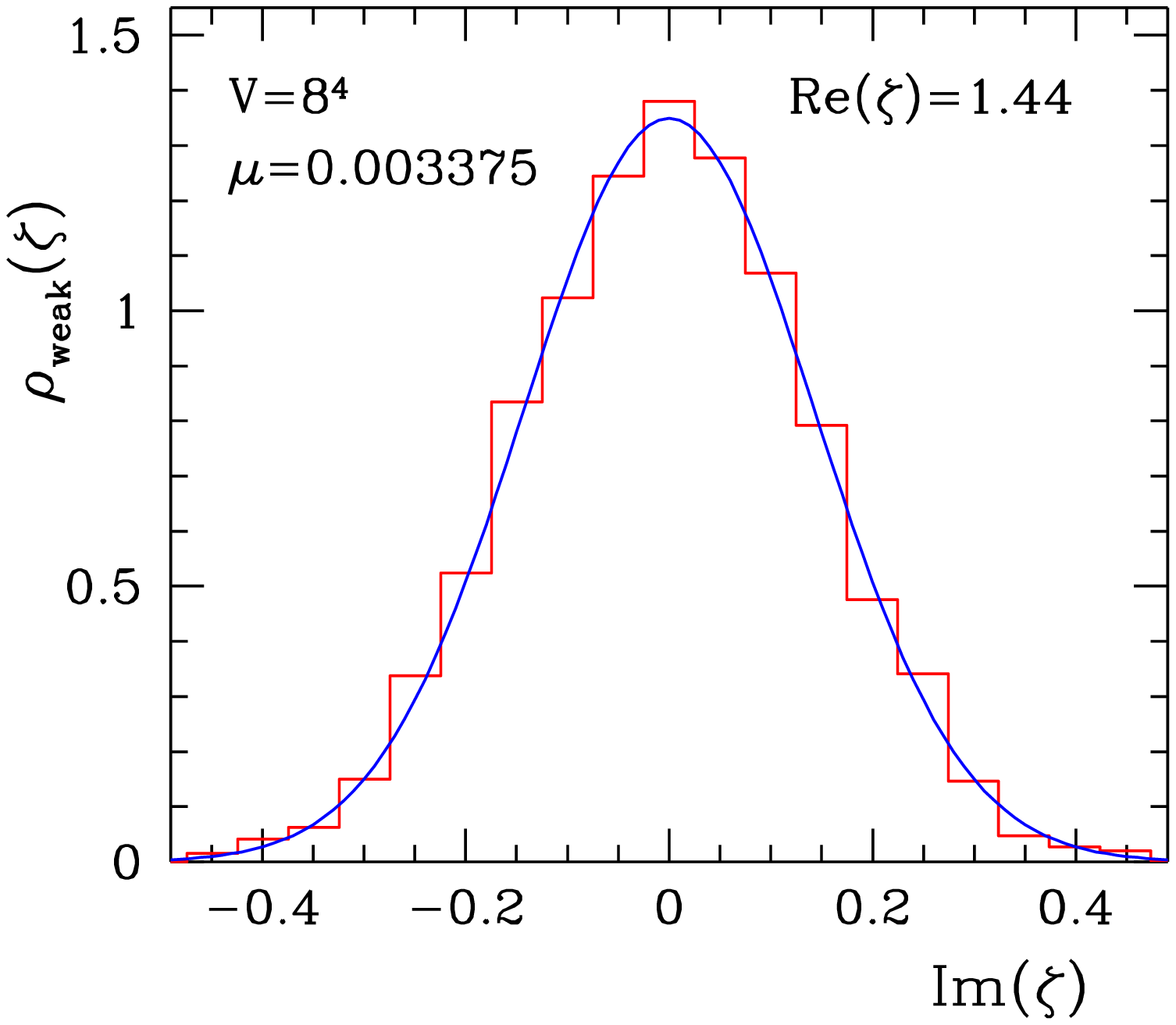,height=33mm} }
\caption{Same as Fig.~\ref{weak6^4} but
for $V=8^4$ and $\mu=0.003375$.
}
\label{weak8^4}
\end{figure}
In Fig.~\ref{weak8^4} we repeat the same analysis for lattice size
$8^4$ and $\mu=0.003375$, chosen to keep $V\mu^2=\alpha^2$ constant in
order to test the scaling predicted by Eq.~\eqref{weaklim}.  Here we
have used the same value of $\alpha=0.20$ and again find excellent
agreement.  This value agrees within $0.3\%$ with the value of
$\alpha$ determined independently by rescaling $\mu^2$ with the level
spacing $d$ from the $8^4$ data.  Similar results are obtained from
the analysis of the $10^4$ data which we do not display here because
of limited statistics.  

Next we turn to the limit of strong non-Hermiticity.  Here, the
eigenvalues are no longer localized close to the real axis and spread
into a two-dimensional domain in the complex plane.  Therefore we have
to modify the determination of the mean level spacing $d$.  Nearest
neighbors are now defined as having the smallest geometric distance
between them.  As can be seen from Table~\ref{table:par}, the level
spacing $d\propto1/\sqrt{V}$ is very different from the weak limit.
The signature of the chiral ensemble \eqref{eq:Zev} compared to the
Ginibre ensemble \cite{Gini65} is a ``hole'' at the origin resulting
from the level repulsion.
(The latter ensemble applies only
in the bulk of the spectrum \cite{MPW99}.)  
To compare the lattice data with Eq.~\eqref{rhostrong}, we rescale the
eigenvalues according to $\xi=c\lambda/d$, where $c$ is a constant
related to the level spacing of our model in the strong non-Hermiticity
limit.  We currently do not have a theoretical result for $c$ and
therefore determined $c=0.82(5)$ by a fit to the data \cite{footnote}.
\begin{figure}[h]
\centerline{
  \epsfig{figure=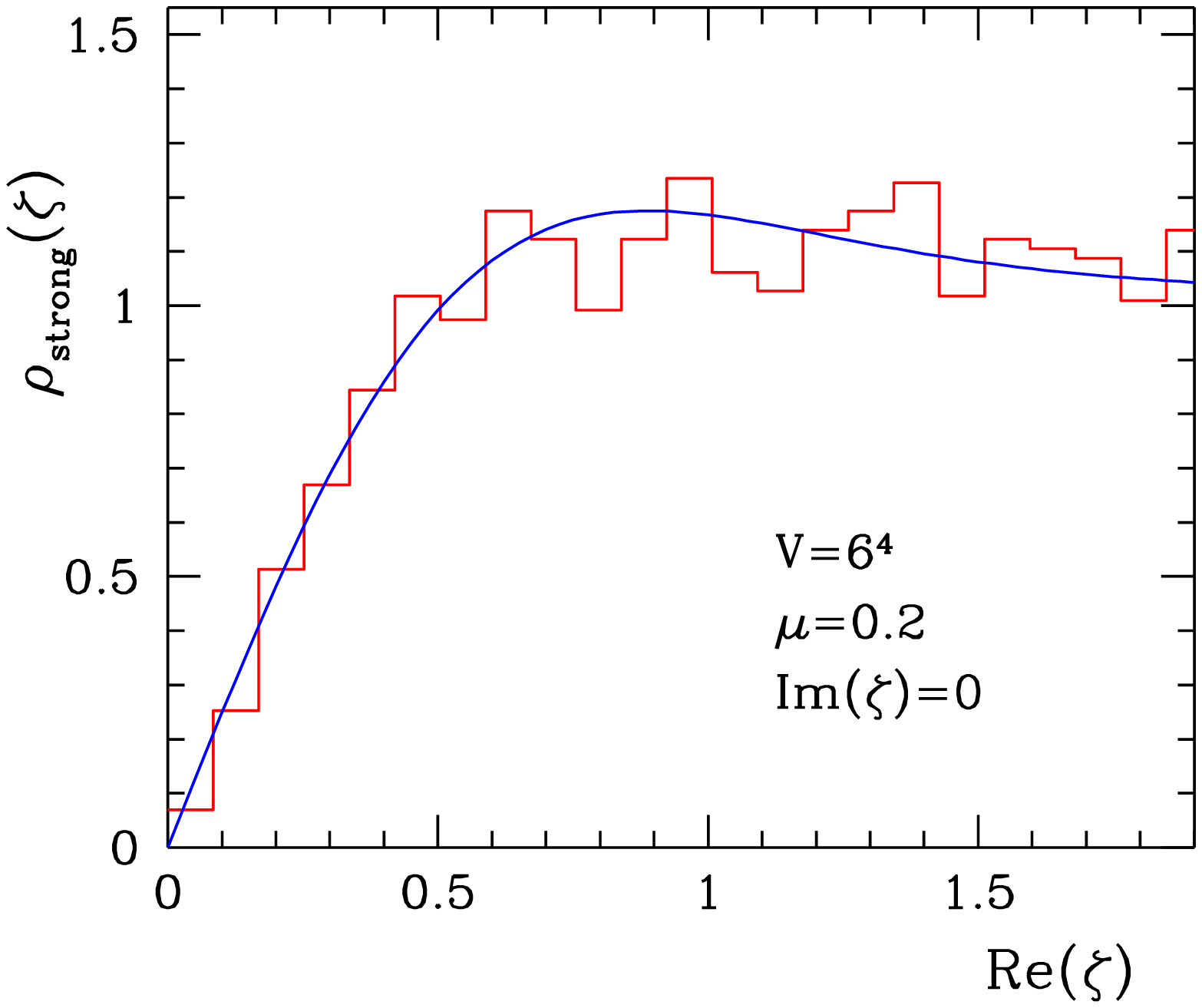,height=33mm}\hspace*{4mm}
  \epsfig{figure=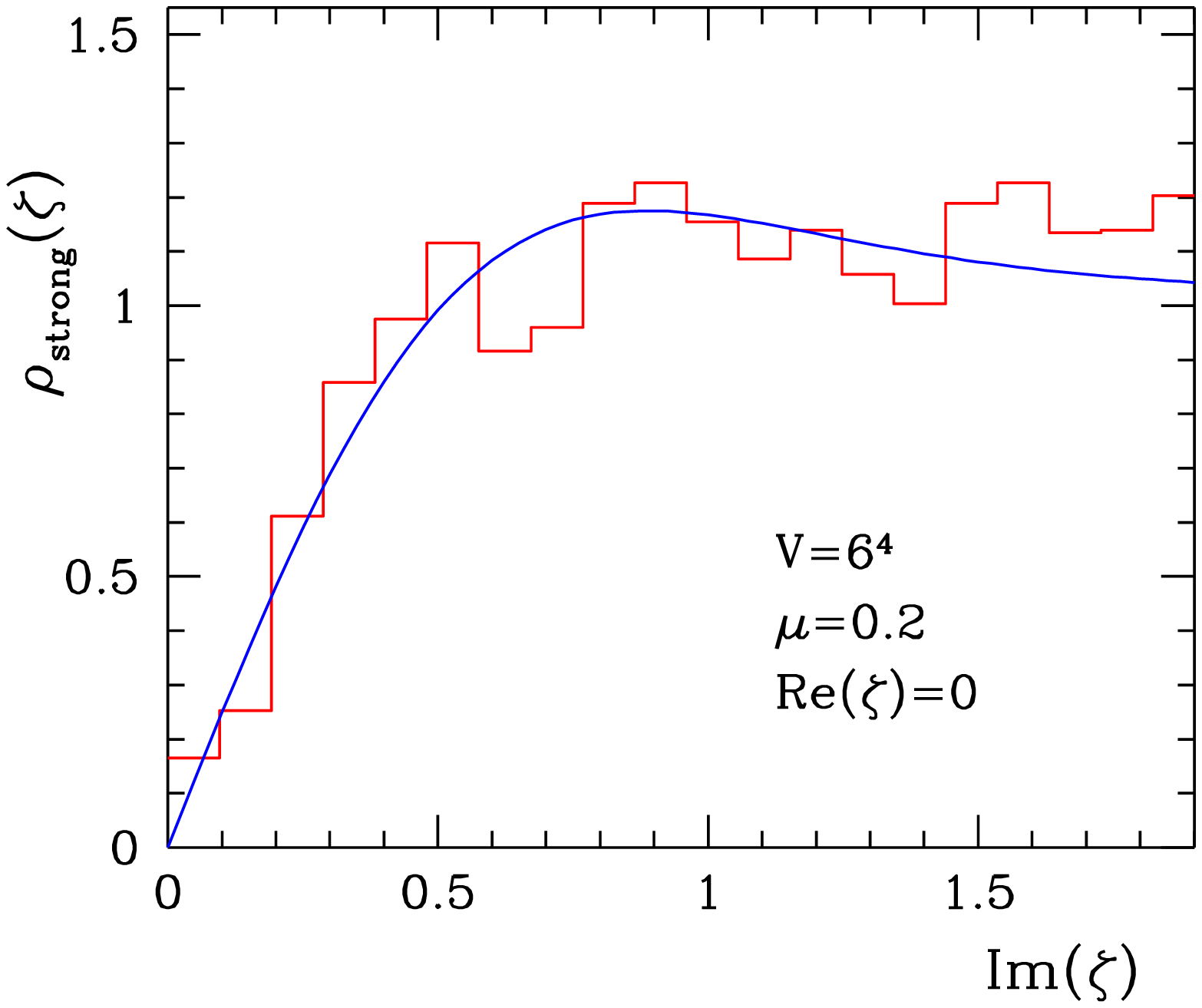,height=33mm}
}
\caption{
Density of small Dirac eigenvalues for $V=6^4$ and $\mu=0.2$,
cut along the real (left) and imaginary 
(right) axes.  The histogram represents lattice data, and the solid
curve is the prediction of Eq.~\eqref{rhostrong}.
}
\vspace*{-3mm}
\label{strong6^4}
\end{figure}
\begin{figure}[b]
\centerline{
  \epsfig{figure=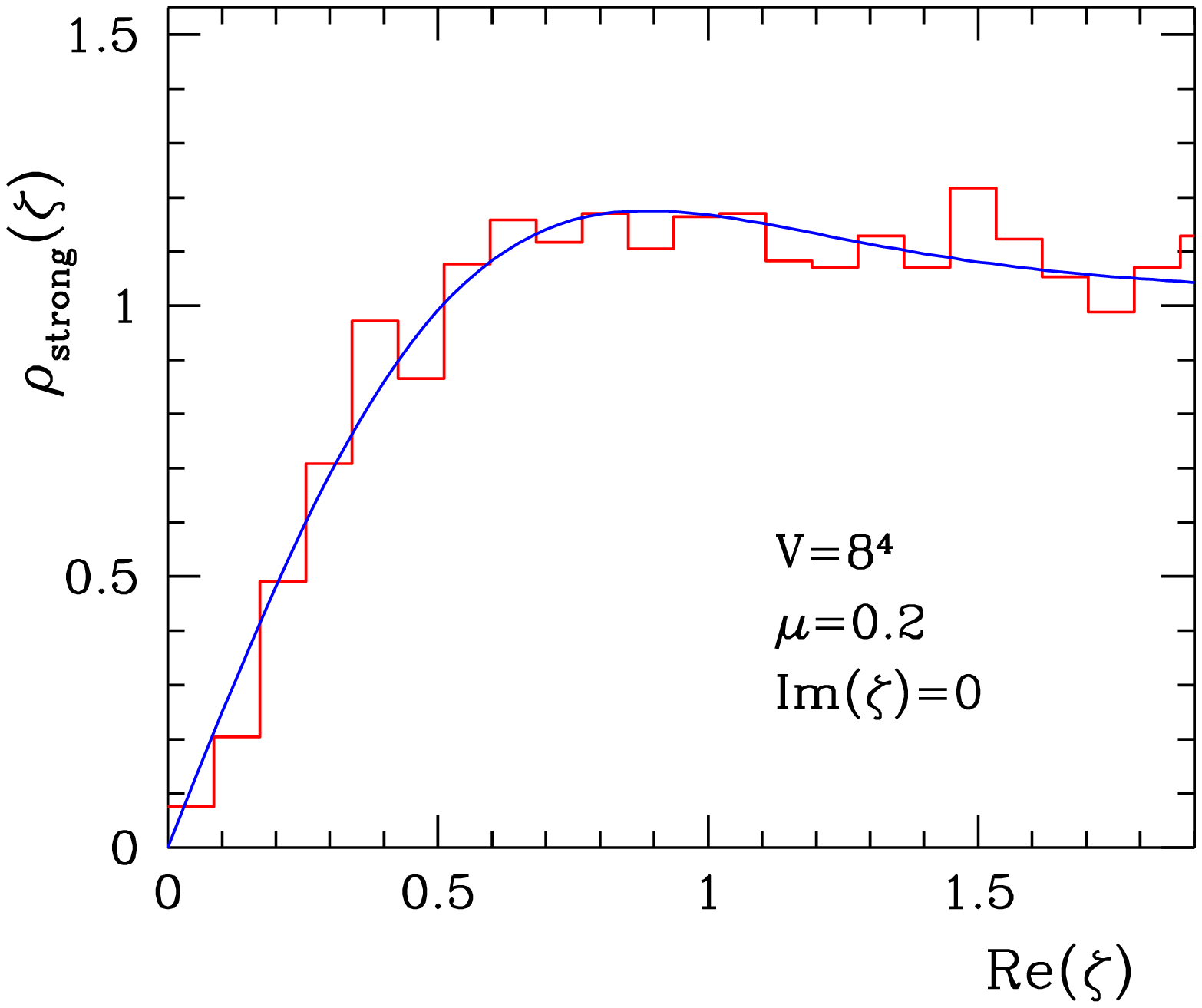,height=33mm}\hspace*{4mm}
  \epsfig{figure=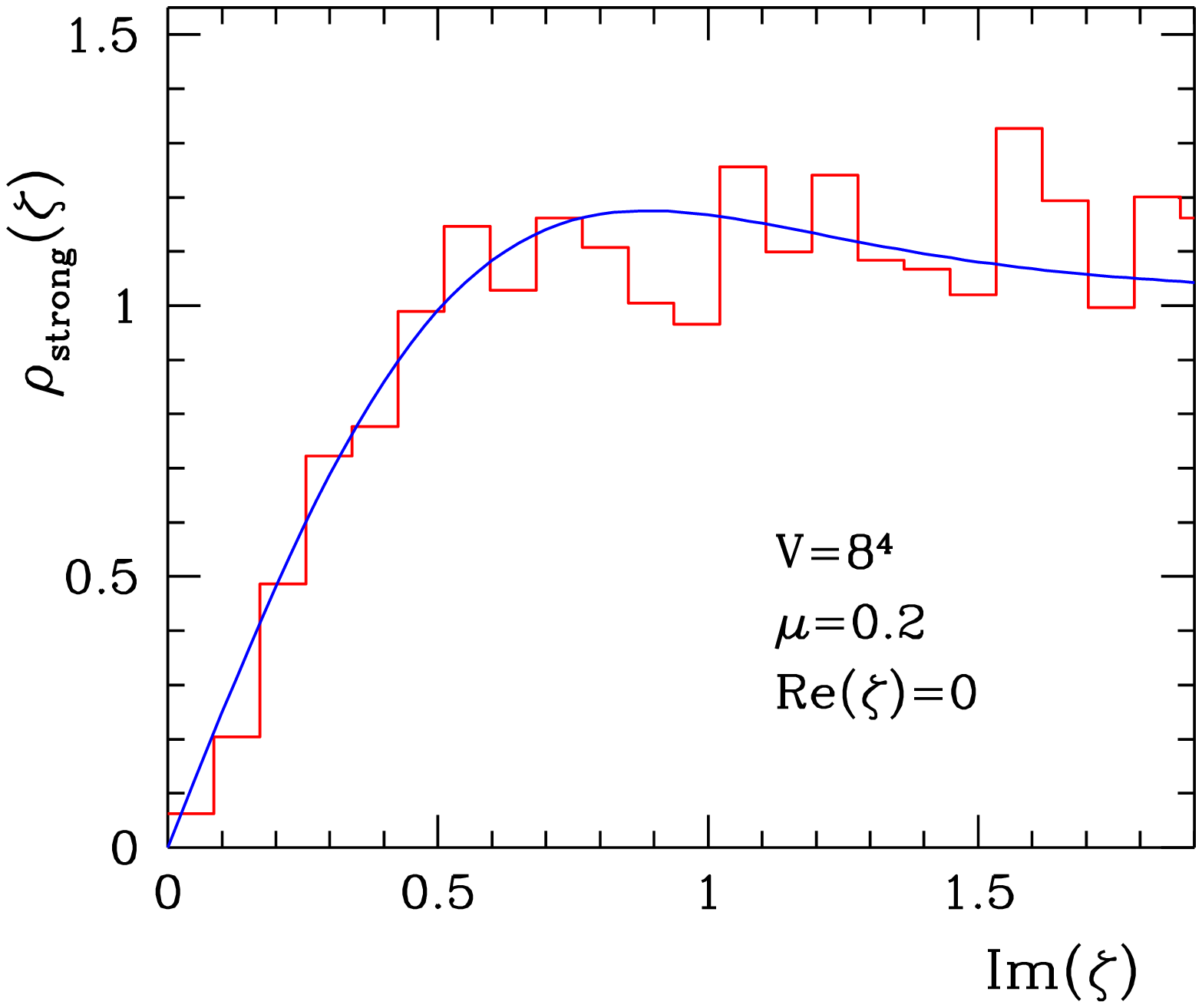,height=33mm}
}
\caption{
Same as Fig.~\ref{strong6^4} but for $V=8^4$ and $\mu=0.2$.
}
\label{strong8^4}
\end{figure}
In Figs.~\ref{strong6^4} and \ref{strong8^4} we plot the data for
$\mu=0.2$ and $V=6^4$, $8^4$ versus the prediction of
Eq.~\eqref{rhostrong}.  Good agreement is found along both the real
and imaginary axis (also for $V=10^4$, not shown), confirming the
rotational invariance of the microscopic density.
Note, however, that there is an important difference between the cuts
in the real and imaginary directions. While the microscopic density,
Eq.~\eqref{rhostrong}, is rotationally invariant, the data spread
macroscopically into a thin ellipse. Along the imaginary axis the
spectrum ends at $\approx \pm3.8$ in our units, while along the real
axis it extends up to $\approx \pm270$.  For that reason the last part
of the histograms along the imaginary axis in Figs.~\ref{strong6^4}
and \ref{strong8^4} may no longer be in the microscopic regime.  In
the data for $V=6^4$ at intermediate $\mu=0.03$, the microscopic and
macroscopic scales no longer separate clearly, and therefore neither
of the Eqs.~\eqref{rhoweak} and \eqref{rhostrong} apply.

In conclusion, we have identified two different regimes in the
behavior of complex eigenvalues of the QCD Dirac operator in the
domain where chiral symmetry is broken.  Our lattice data confirm the
predictions of random matrix theory quantitatively, both at weak and
at strong non-Hermiticity.
Matrix models
thus provide a detailed theoretical understanding of the properties of
complex Dirac eigenvalues in a new regime with non-vanishing
chemical potential, including the correct scaling of eigenvalues and
chemical potential with the volume.  Our findings may have
algorithmical implications, since it is typically the low-lying Dirac
eigenvalues which determine the numerical effort in lattice
simulations.  We wish to emphasize that, although our conclusions are
based on quenched simulations using staggered fermions, the
predictions of the matrix model could also be tested in unquenched
simulations and in sectors of nontrivial topological charge.
This will be the subject of future work.

This work was supported by the DFG (G.A.) and in part by DOE grant No.
DE-FG02-91ER40608 (T.W.).

{\it Note added in proof.}---The microscopic density of the model
Eq.~\eqref{Zsteph} was recently computed in the weak non-Hermiticity
limit \cite{SV03}.  The result deviates very slightly from our
Eq.~\eqref{rhoweak}, but this difference cannot be resolved by our
data.

\pagebreak

\pagestyle{empty}
\onecolumngrid
\begin{center}
  {\large\bf Erratum: QCD Dirac Operator at Nonzero Chemical Potential:}\\[1mm]
  {\large\bf Lattice Data and Matrix Model}\\[1mm]
  {\large\bf [Phys. Rev. Lett. 92, 102002 (2004)]}\\[5mm]
  Gernot Akemann and Tilo Wettig\\
  (Received 8 December 2005; published 18 January 2006)\\[1mm]
\end{center}


In Figs.~1 and 2 we compared lattice data to the analytical
prediction of Eq.~(5), which pertains to the weak non-Hermiticity
limit of the model in Eq.~(1).  For such a comparison, two independent
parameters must be determined: the mean level spacing $d$ and the
dimensionless parameter $\alpha$ appearing in Eq.~(5).  These two
parameters can be expressed in terms of two low-energy constants, the
chiral condensate $\Sigma$ at $\mu=0$ and the pion decay constant
$F_\pi$, by the relations $d=\pi/\Sigma V$ and
$\alpha^2=2\mu^2F_\pi^2V$ \cite{AOSV}.  Here, $V$ is the physical
volume.  We erroneously determined $\alpha^2$ as $(\mu a)^2\pi/\sqrt2
da$, where $a$ is the lattice spacing.  This amounts to setting
$(F_\pi a)^2/\Sigma a^3=2^{-3/2}$.  Instead, $F_\pi$ and thus $\alpha$
should be obtained either by an independent measurement of $F_\pi$ or
by a fit of the data to Eq.~(5).  Performing the latter for the data
sets in Figs.~1 and 2 yields a value of $\alpha=0.200(2)$, which is
consistent with the value of $\alpha=0.20$ used in the Letter.  Thus,
Figs.~1 and 2 as well as the conclusions of the Letter remain
unchanged.

The fitting of $\alpha$ to the lattice data constitutes a new method
for obtaining $F_\pi$ through the relation quoted above, see also
Ref.~\cite{OW}, but we refrain from applying this method here since
the lattice data were obtained in the strong-coupling region.  A
related method to determine $F_\pi$ using isospin chemical potential
was introduced recently in Ref.~\cite{DHSS}.

We thank J.C. Osborn for pointing out the error described above and
P.H. Damgaard for useful discussions.


\begin{thebibliography}{99}
\bibitem{footnote} We have checked for several combinations of $V$ and
  $\mu$ that we obtain the same value for $c$.
\end{thebibliography}
\end{document}